\pdfoutput=1
\documentclass[12pt,1p]{elsarticle}%
\usepackage{amsfonts}
\usepackage{amsmath}
\usepackage{amssymb}
\makeatletter
\def\ps@pprintTitle{%
	\let\@oddhead\@empty
	\let\@evenhead\@empty
	\let\@oddfoot\@empty
	\def\@oddfoot{\reset@font\hfil\thepage\hfil}\def\@oddfoot{\reset@font\hfil\thepage\hfil}
	\let\@evenfoot\@oddfoot	
}
\makeatother
\providecommand{\U}[1]{\protect\rule{.1in}{.1in}}

\begin{document}

\begin{frontmatter}%


%
\title
{Physical Reality and the Unobservables of Physical Nature \footnote{Published in {\bf Einstein, Tagore, and the Nature of Reality}, P. Ghose (Ed.), Routledge Studies in the Philosophy of Mathematics and Physics (2017).}}

\author{C. S. Unnikrishnan}%

\address
{Tata Institute of Fundamental Research, Homi Bhabha Road, Mumbai 400005, India\\
E-mail address: unni@tifr.res.in}%

\begin{abstract}%

The fundamental physical theories that interpret and explain behaviour of matter in nature are dependent on several unobservables and insensibles in their construction. While a rigorous natural philosophy cannot take them for granted, there does not seem to be a way of avoiding such unobservables in our theories.  While a program to banish all unobservables from physical theory is unlikely to succeed, and perhaps even unnecessary, they are both the strong and weak points of the theoretical descriptions of physical nature.  Analyzing them for empirical and philosophical consistency and integrity is always a promising path towards a better theory. In this paper, I examine the nature of physical reality in the context of unobservables in physics and discuss three examples. One is about the apparent loss of physical reality due to the need for a consistent quantum mechanical representation. The second example deals with the conflict between the assumed reality of  quantum fields, so fundamental and essential to our standard physics worldview,  and the dynamics of the observable universe.  The third deals with an all-important difference between conventional modern physics constructed in the unreal and unobservable empty `space'  and an empirically and logically determined physics with matter-filled universe as its arena. The acknowledgment of the observable matter-filled universe necessitates reformulation of dynamics with total relativity. Not surprisingly, this paradigm with its universal cosmic links also holds human concepts of harmony and beauty.

\end{abstract}%


\end{frontmatter}%



\section{Introduction and Scope}
The fundamental physical theories that interpret and explain behaviour of matter in nature are dependent on several unobservables and insensibles in their construction. Entities like fields, wave-functions, and even space and time are all unobservables, except as manifestations of material existence and behaviour.  There is thus an obvious difference of degree and meaning between the reality associated with these unobservable theoretical entities and that of perceptible matter.  The success of the physical theory is often taken as evidence for the physical reality of such unobservables.  While a rigorous natural philosophy will not be able to support or approve their reality with the same vigour and conviction as it might defend the reality of matter, there does not seem to be a way of avoiding such unobservables if we have to construct theories. Though there is compatibility and consistency between observables and unobservables in most of classical physics, apparent conflicts and dissonance arise when microscopic physics is to be understood with a satisfactory theory. There are even observational consequences highlighting such conflicts when cosmology and the dynamics of the universe are included into the larger physical framework.

In this paper, I examine the nature of physical reality in the context of unavoidable unobservables in physics \cite{Unni-Beyond,Unni-Bili},  and discuss some examples. For the purpose of this discussion I will work with a definition of an unobservable (in physics) as a quantitative entity, created and mathematically representable in relation to a theory of sensible matter and its behaviour,  but whose ontological status cannot be established nor demonstrated directly or by deduction employing methods usually used for material entities. Thus an unobservable cannot be proved to exist by tangible sensing with any conceivable observational device or even by a logical argument relying on empirical evidence, with a degree of conviction nearing that of the reality of material existence that we normally admit in the context of physical science. Another description, in line with the history of the use of such unobservables, would be as entities created or postulated, in the context of the physical theory, which we are willing to believe as real but cannot be proved to exist or otherwise by any empirical means known to science. In this sense, such unobservables belong to a projected reality involving the human mind. Clearly, we do not have a firm definition to go by and I hope that the examples can serve as clear indicators. 

I do not plan to discuss the issue of realism in the sense it is usually debated on, where questions or doubts are raised whether what is perceived implies an underlying objective reality and existence. For the purpose of our discussion such debate is a distraction. In fact, the view taken here is that the issue of real existence independent of perception of the human being is irrelevant for human endeavours like science. All what is required for science is the internal consistency and stability of perceived and observed patterns of nature.  While it is a reasonable to extrapolate that to an objective reality underlying perceived phenomena, there is no logical necessity that we establish that reality before building theories of phenomena. However, this is not a point to dwell on within the scope of this paper.  

One of the features of science, especially physical science, is a powerful urge to construct theories of phenomena, instead of just cataloging and classifying phenomena and their spatio-temporal patterns. Theories go beyond mathematical modeling of patterns.  For example, the mathematical and geometrical description of planetary motions, which of course contains some aspects of theoretical construction of a mathematical nature, attains the status of a physical theory only when completed with an explanatory description involving the massive sun, gravity, inverse square law etc.  The ability of the theory to include a reasonable explanatory power, addressing the questions of `why' and not just `how' is important in natural philosophy and physics. 

As soon as we embark on this program of theoretical description of observed phenomena, we face the problem of having to conceive and include entities that are not directly sensed, but could reasonably be argued to `exist' in nature. More often than not, such entities remain unobservable, with only plausibility and consistency arguments in support of their reality. However, they are, by design, fully representable mathematically and they bear definite relations to observable quantities and other elements in the theory.  Often, the motivation for introducing an unobservable is a strong commitment to causal development of the physical world through a series of cause-effect relations. The justification is made stronger subsequently if there is relatively long lasting success and consistency of demonstrating the cause-effect relations in phenomena relying on such an entity.  A concrete example, the theory of planetary motions, can clarify the essence of such a structure of the physical theory.  

\section{Physical Theory with an Unobservable: an Example}
There were pre-Keplerian descriptions of planetary motions with supporting geometrical construction that did well in the construction and prediction of the ephemeris. Kepler's model of planetary motions, based on a compact set of statements or `laws' improved the precision of dealing with planetary trajectories, but still did not provide an answer to why planets move in elliptical orbits in precisely the way they do. That explanation had to wait for Newton's theory of gravitation and his identifying the long range gravitational action as responsible for holding the planets in their orbits. All of Kepler's laws emerged as consequences of  the underlying physical theory advanced by Newton, with the inverse-square law for gravity and the conservation of angular momentum. But the transition from mere laws to a physical theory also brought in the need to hypothesize the phenomenon of non-contact gravitational action that can `hold' an object at a distance under a force without any contact. The disbelief about such a possibility is cured by postulating a field of gravity (which was done much after Newton's time), continuous and contiguous from one material body to another, acting like an invisible and insensible, yet `real' entity. We clearly see in this example the necessity and role of an unobservable in a fundamental physical theory.  Faraday is rightly credited with firmly presenting the concept of the ethereal fields of electricity and magnetism, and with asserting and defending their `reality'. Phenomena based on electromagnetic radiation are now considered sufficient proof for their physical reality. Any attempt to avoid such pervading fields makes the theory very complicated and subject to criticisms of non-local action at a distance. In any case, historically, a `field' has become an integral part of physical theories, in spite of our inability to really prove its reality with the same degree of conviction as the reality of matter.

\section{Fields, Sources, Space and Time} 
The concept of `fields' of various kinds dominates  the physical theory and modern physics, and fields are essential to both classical and quantum theories. However, the classical fields we are familiar with are somehow in good harmony with our notions of plausible reality, perhaps because such fields were postulated in an intuitive way in the context of electromagnetism and gravity. They are in fact natural extensions of sensible material fields, like the velocity or density at different points of a flowing liquid, or the temperature at different locations in space. These material fields are related to some physical property of continuously distributed matter. Hence, once the reality of matter is accepted, the reality of such a field is not in serious doubt. 
In contrast, a magnetic field or a gravitational field is characteristically different. They certainly need a material source, like a current or a massive body, but fields are posited to exist in regions where there is no source. The field is the physical device that connects one material object to another spatially separated one,  and the field is the agent of the interaction and the evolution from cause to effect. Such fields are `observed' only through the dynamical behaviour of a material body - a test particle, as it is often described - in the presence of some spatially separated source matter. Here we see a secondary layer of reality that can be questioned into doubt. Fields like the electric and magnetic fields, and the gravitational field are theoretical constructs, posited to exist to simplify both the mathematical structure of the theory and to preserve an intuitive understanding of physical phenomena as due to contiguous cause-effects relations in space and time. One of the implied tenets of physical sciences is the need to empirically prove the existence of an entity as `physically' real. However, this cannot be insisted on in the case of such fields.  They remain unobservable in the sense of `observability' as applied to the reality of matter, while being consistent with the hypothesis of their reality in physical phenomena.

Einstein had discussed the relation between matter, fields and space in the context of the development of the theories of relativity \cite{Ein-Franklin,Ein-Worldview}. The  point of view expressed there is that fields are physical states of space, modifiable by the presence of material sources. (In fact, this was his motivation for constructing unified field theories where there is only one structure of space when both gravity and electromagnetism are considered together). Though he gave importance to sensible experience as the true basis for concepts with ontological content, one sees in his writings the preference to give superior status, relative to matter, to the geometric empty space. (In his article  `Physics and Reality' (ref. \cite{Ein-Franklin}) he compared his field equations to a building with one of its wings made of fine marble, standing for geometry of space-time, and the other made of low grade wood, meaning matter. That he had to depend on two forms of matter itself to make this statement is another matter!).

There is a strong relation between the familiar fields and the state of motion of matter. While an electric field represents the action between static charges, the magnetic field represents the action of charges in motion, or `currents'. When motion ceases, magnetic fields disappear. Also, curiously, only moving charges can `see' a magnetic field.   There are similar features in the modern theory of gravitation. Hence, the reality of such fields is dependent on the state of the motion of the observer, or that of the device used by the observer to `sense' the field. This feature illustrates the subtle aspects of dealing with the issue of reality of even familiar entities in the physical world.

At least in one context, the hypothesis of a `field' seems to be more than a convenience. That is the phenomenon of radiation. Light can in principle be thought of as the response of charged particles in the eye to another charged particle moving periodically somewhere else far away. In this view, there are only (charged) material particles. However, thinking of light as a propagating field in its own right and reality, separated from the source or a potential detector, is considered to be essential for a consistent description. Indeed, light of that kind is considered as particles in their own right, called photons. Therefore, understanding the reality of photons is, and will remain, crucial for understanding a whole lot of ontology in physical theory. One important aspect to note in the case of such `radiation fields' is that they are retarded in time - there is a time delay that increases with distance between the cause (motion of the source) and the effect (response of a detector). So, perception itself is not perception of `now' in an absolute sense. This is the basis of the denial of absolute simultaneity of events in relativity theories. We will come back to a discussion of the field of electromagnetism, when we discuss the  `physical properties' of the unobservable vacuum or emptiness.   

There are two fundamental `fields' used in physics, with progressively more physical roles ascribed to them as the theory progressed. These are space and time. While Newtonian physics needed space and time with some precision in their definition distilled from the common sense use of these concepts, modern relativistic physics treats these as dynamical fields in a sense similar to the way electromagnetic fields are treated. A fuller discussion of this is not intended here. However, we will occasionally comment on the status of space and time, or space-time as it is referred to in relativistic physics, as physical fields. A discussion of the reality of space and time is fraught with many dangers of logical and philosophical complexities. Simply put, a concept like time does not make sense without matter. All our notions of time are changes in material configurations. While space without matter is `imaginable' in a way, even those notions of spatial extent etc. are generalized from our familiarity with material reality. Quoting from Einstein's forward to Max Jammer’s book \cite{Jammer},  `Concepts of Space',
 \begin{quote}
 	...two concepts of space may be contrasted as follows: (a) space as positional quantity of the world of material objects; (b) space as container of all material objects. In case (a), space without a material object is inconceivable. In case (b), a material object can only be conceived as existing in space; space then appears as a reality which in a certain sense is superior to the material world. Both space concepts are free creations of the human imagination, means devised for easier comprehension of our sense experience.
 \end{quote}

Space and time are the supreme and primary unobservables in physics, without any reality independent of matter, notwithstanding their modern status as the dynamical arena of gravitational effects as described in Einstein's general theory of relativity. However, they are essential for even starting to imagine about a physical theory, let alone construct it. The situation is different from the notions of electromagnetic fields or the gravitational field because space and time are continua without material sources. Their existence and reality are the most primitively entrenched notions in our minds and yet they are primary examples of unobservables in physics. The difficulty of incorporating space and time into the physical theory without referring to matter has been realized from the very early times of mathematically precise physical theory as evident in Newton's Principia \cite{Principia}: 

\begin{quote}
Absolute space, in its own nature, without relation to anything external remains always similar and immovable. Relative space is some movable dimension or measure of the absolute spaces, which our senses determine by its position to bodies, and which is commonly taken for immovable space.
But because, the parts of space cannot be seen, or distinguished from one another by our senses, therefore in their stead we use sensible measures of them. For from the positions and distances of things from anybody considered as immovable, we define all places, and then with respect to such places we estimate all motions…

\end{quote}

As it is well known, a well reasoned critical thesis on space without material references as the arena of physical effects (like the inertial forces) had to wait another 200 years, till Mach's critique on Newton \cite{Mach}.  Unfortunately, modern physics went retrograde on this aspect, denying Machian insights; we will see later that we are destined to pay the heavy price for this. 

\section{Quantum Physics and its Unobservables}
Twentieth century physics brought in new notions of particles, trajectories and causal relations for dynamics (also called `laws' of physics), which dissolved or even denied the earlier established notions of identity, individuality, distinguishability and localizability in space and time – hence a new notion of reality was to be developed.

Most of modern discussions on reality in physics are linked to the microscopic physics described by the quantum theory or quantum mechanics (QM) \cite{Espagnat-Reality}. A major instigator for this situation was one of the prominent contributors to the theory, Einstein himself, who asked the most pertinent questions about the representation of physical reality in QM \cite{Pais}.  It is in the context of QM that one first encounters the necessity to represent the dynamics of a particle with an entity that has properties that are familiar from classical physics, but has no tangible existence in real space - a wave to which properties like wavelength can be ascribed in direct relation to mechanical properties of the particle, but has no real existence as a physical wave that propagates in real space. The `wave-function' (or the $\psi$-function, as Einstein and Schr\"odinger preferred) in QM is an unobservable field without a source.  It is inseparable from the existence of the material particle it represents, yet it is not the material particle. It holds in its description all the physical properties of the material particle, like charge, mass etc., and interacts with other material sources in a conventional way through their `fields', but it has no tangible traces in real space. It can be split into multiple parts by simply providing possibilities of splitting, with each part holding physical properties of the material particle it represents in entirety, yet the different parts do not interact with each other in any way. In short, no consistent real physical existence can be ascribed to a wave-function. However, the entire description of microscopic physics today depends on the wave-function representing all possible physical behaviour of the particles as if it exists in some way in space, feeling the external world through interactions. 

The only tangible relation between the wave-function and observable entities is statistical - the absolute square of the wave-function is related to the probabilities of relevant observations. Every observation in general `resets' the wave-function, so to speak, and a new evolution starts. Perhaps the QM wave-function is the only unobservable in physical theory that has not been assimilated into the common sense of the physicist. 

It is tempting to commit the mistake of identifying the wave-function with the particle in many situations, loosely calling it a `matter-wave', especially in those situations where one deals with just one particle. The underlying representational feature is the familiar wave-particle duality. Indeed, this was the kind of intuition that people tried to cultivate and discuss in the early days of QM. This program fails in details and no tangible reality or even a describable nature of existence could be assigned to the wave-function.  However, even today many discussions, especially those dealing with quantum interferometry involving particles (called matter-wave interferometry), try to describe the underlying physical phenomena as if the wave-function has a real physical existence in space, which is generally an inaccurate and inconsistent notion. 

\subsection{Some problems arising from the unobservable of QM} 
Before we discuss some problems that arise from the use of the unobservable wave-functions in QM, I want to stress that there is no known conflict or inconsistency between QM and the assertion that material objects have an objective reality of existence, in the sense of   their possessing some physical properties as well as occupying some region of space even before an act of observation has been done. Indeed, writing a wave-function in QM assumes the physical reality of the material system with some physical properties, like mass, spin etc.  Therefore, commonly found statements like `an atom or even a stone has no properties prior to observation', or `the moon exists only when it is observed...'  \cite{moon} etc. are not implications of QM. However, what is true in the QM representation employing a wave-function is that the representation cannot be directly interpreted as providing a correspondence to reality in any familiar terms. Thus, the wave-function does not hold in itself information of exact position, exact velocity, exact components of angular momentum etc. The wave-function represents the physical quantities without `possessing' the physical properties of the particle, like energy or momentum. Yet, it holds the exact probabilities for these quantities taking specific values or ranges of values on observation. Also, the wave-function is a divisible entity, much like a real wave, whereas the underlying material system is not. What is observed or sensed is only the material system, and then the wave-function transforms suitably to represent the new information of observation, and many of its multiple parts disappear from consideration. No one is competent to say whether it disappears as a physical entity because its ontology is not known yet. The correspondence between matter and wave-function in the theory is such that only the probabilities of realization of particular values of physical quantities are transformed or renewed, and the physical system and the totality of its properties (like total energy) remains conserved whereas the wave-function itself is not.  There are also situations where there is a QM representation of the physical properties of a material system consisting of many (possibly two) particles, but there is no QM description at all for any of the particles taken individually.  Therefore the ontological status of the wave-function and its reality as a physical entity etc. are  not properly understood.  

Yet, there is something physically real about the wave-function as can be seen from a simple consideration. Figure 1 represents a situation of 'interference', familiar in the case of light or other waves.  Essentially, the amount of light that exits from port 1 or 2 depends on the physical properties, like refractive index and path-length, of both paths, A and B. Variation of the difference in path lengths changes periodically the amount of light from either port between a maximum and zero (or a minimum close to zero). 

\begin{figure}
	\centering
	\includegraphics[width=0.7\linewidth]{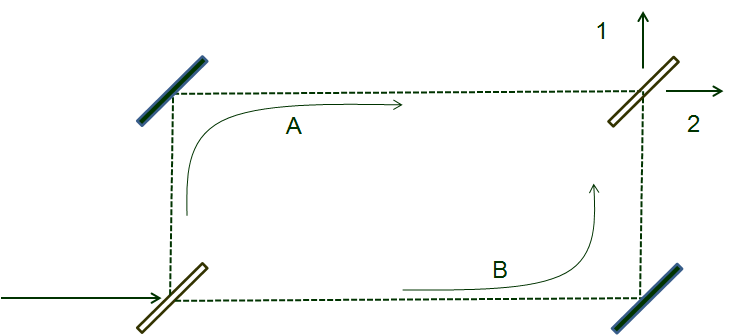}
	\caption{Quantum interference depends on the details of both paths in space and time and demands a space-time ontology and interpretation for wave-functions as well.}
	\label{fig:figure1}
\end{figure}

Therefore, one can have a situation where the entire light exits one of the ports with the other remaining totally dark. Small changes in the path length then can change the situation, and reverse it. The bizarre possibility in QM is that similar interference happens for material particles sent in one by one into such a device,  as if each particle has an associated wave \cite{Ghose-QMtests}. Irrespective of what the physical picture is, the fact remains that the port through which the particle exits can be controlled by small changes in either or both of the possible paths and the total change from port 1 to port 2 can be done deterministically by changing one or both of the path-lengths, by a physical movement of material elements like the mirrors. It is then very difficult to maintain the view that there was nothing physically real in those paths even though moving some mirrors in those paths changed a physical outcome. If one accepts that physical changes could happen only because the mirror interacted with and affected something real in space, then one will also have to accept that the entity was present in both paths simultaneously, even though the material particle could not exist in both paths simultaneously. Further, one will have to face the consequence that if the particle was looked for and found in one of the paths before the exit ports, then the wave-function corresponding to the path where the particle was not found has to disappear from the theory and by implication, from any possible physical existence--that step is required to preserve the relation of the wave-function to probabilities of observations, called the Born's rule. 

\subsection{From classical reality to quantum unreality}
A classical particle sourced with some random energy E will have some energy, momentum etc., the definite values of which can be known only after a measurement. But we do not contest that the particle possesses some specific energy even before measurement, which shows up when an actual observation is made. So, the energy state of the particle might be specified as ($E_a$ or $E_b$  or...$E_s$) with corresponding probabilities ($p_a$, $p_b$...,$p_s$). In this case, we believe that a particle does possess one of these energies between the source and the detection, even though we specified the physical states with a possible set of energies and probabilities. In QM, the physical state is not specified like this though this classical statistical feature can also be included in the specification. In the simplest cases of dealing with a particle, it may be that the state is specified through a wave-function as $\left|S\right> = a\left|S(E_a)\right> + b\left|S(E_b)\right> +...+ s\left|S(E_s)\right>$, called a superposition of states with different energies. In this case, the square of each term (a,b,...s) encodes also the probability of observing the particle with that particular energy. However, it will be wrong to interpret that the particle has one of these energies before an observation. It is also wrong to say that it had all these energies at the same time. Similarly, if a particle is actually observed in a small region of space, QM does not allow any claim that it was somewhere nearby just before the observation.  The '+' sign here does not correspond to the 'or' in the specification of the classical state. Neither does it represent our lack of the precise knowledge of the physical state. Any translation of the mathematical state into any other language does not seem to capture its meaning. \emph{The unobservable in the theory is also untranslatable}. This is perhaps the point in the quantum description where the familiar concepts of reality lose their anchor. The best one might do is to say that each term $\left|S(E_s)\right>$ represents some kind of abstract wave with energy $E_s$, and we have a superposition of such waves as the representation. But this does not help much because the observations are on the material particle and will result in some specific value of energy, and then we are forced to say that all other waves in the superposition simply disappeared after the observation. In spite of this, the particle has a definite QM physical state, $\left|S\right>$. There is no confusion on this fact. So, within QM, a physical state need not be restricted to a state with specific values of each physical quantity. There is a well-defined QM state even though we are not able to answer questions like what is the energy of the particle.

The main features of the QM representation can be discussed in a simple manner. The single characteristic feature is superposition of the representations of physical states and the resulting `incompatibility' or mutual uncertainty relation between measurements of some pairs of quantities.   There are many such `incompatible' physical quantities in the quantum world, and examples are position and momentum, spin in the x-direction and spin in the y-direction etc.  In fact, this incompatibility between certain pairs of observable properties, contained in the possibility of superposition, is what distinguishes QM from classical physics. 

If a physical system has a measurable property A, perhaps with just two possibilities of outcomes labeled H and L, QM represents this as two possible `states of being' (H) and (L) with associated measurement values, and asserts that any linear combination of the states (H) and (L) is also a valid and possible state. Obviously, such a linear combination, say (H)+(L) = (U), cannot give a definite value H on all measurements. (Here we have ignored writing the numerical factors because it is not relevant for the discussion.  However, they are important because the square of the coefficient is the probability to find the system in that particular state, if observed).  If the superposition is in equal proportion, the theory associates equal probabilities for observation of H and L and in general the square of the coefficients of the superposition determines the associated probabilities. 

Consider the example of an atom, which can either be in the state `High' (H) or `Low' (L). Hence a possible QM state of the atom could be written as the superposition $(H) + (L)$.  Though one might be perturbed about the meaning of the state of being like $(H) + (L)$, once it is understood that it is a representation with no implication that the atom is both High and Low at the same time, one can progress further. All we need to accept at this stage is that the state $(H) + (L)$ represents a valid physical state, even if we are not able to immediately `understand' what it means. In some cases, there is a straightforward interpretation for such states which is comforting. For example, if the state labeled $(+Z)$ represents a state of the spin of an electron being along the Z direction (in x,y,z coordinate system) and the state $(-Z)$ represents it being in the negative z-direction, the state  $(+Z) + (-Z)$ is in fact the state in which the spin is along the x-direction.  In the hypothetical case we were discussing, the state $(H)+(L)$ might be a definite state for some other property, possibly dichotic,  with symbolic values `Up' (U) and `Down' (D).  The scope of QM is contained in the representation that allows superposition where combinations of two states of one property express also two states of some other property.  For example, symbolically, the two states (U) and (D) can be expressed as superposition of (H) and (L): $(U) = (H) + (L)$ and $(D) = (H)-(L)$. Adding complex numbers to the scenario allows more measurable properties and states to be expressed. 

This simple representational feature contains the essence of QM reality. We see the uncertainty principle in action right away. If the physical system is prepared in the physical state (U), for example, but we decide to observe with a suitable apparatus to see whether the system is in state (H) or (L), we will get random results with equal probability, because the state (U) is also the state $(H) + (L)$. Also, a state prepared as (H) is really the combination $(U) + (D)$ (with a multiplying numerical factor which we ignored) and the uncertainty is reciprocal. Clearly, QM uncertainty is not because the source sends out randomly the states (U) and (D) or (H) and (L). The representation of physical states with superposition precludes the naive reduction of QM to a classical statistical theory.

QM explicitly prohibits any interpretation that if the value H is found on observation, then the state was necessarily (H) just a moment before the observation. It could have been (U) or (L) or some more general superposition of the two. That is the crucial difference between classical statistical observations and the ones in QM. Therefore, in QM, there is a collapse of the state on observation; a sudden change from a general state of superposition to a reduced state. This is not a sudden change in just our knowledge of the physical state of the system (as it would happen in classical physics as well), but it is a change of the physical state itself, within the QM formulation. 

Unfortunately no real progress has been made on clarifying the associated ontology, or the lack of it, throughout the entire history of QM, now nearing a century. If at all, the puzzles and surprise have become stronger due to the many interesting experiments that have been performed \cite{Ghose-QMtests}. 

\subsection{The case of two particles and two properties}
The representation of the physical state involving two particles, each possessing two physical properties that cannot be simultaneously specified to arbitrary accuracy, reveals some subtle aspects of reality in the context of quantum mechanical description of the microscopic world.  It is this kind of an example (with position and momentum as the physical variables) that Einstein and collaborators discussed in the EPR (Einstein, Podolsky and Rosen) argument in 1935 that the QM representation is not a faithful representation of the real physical state \cite{EPR}.  Almost the entire modern discussion on reality in the quantum world is in the context of such examples and the EPR argument.  

Given the possibility of superposition, the joint state of two atoms could be (H)+(L) for each in general and then we can write the joint state algebraically, putting labels 1 and 2 for the particles, as
\begin{equation}\label{key}
[(H) + (L)]_1 [(H) + (L)]_2 = (H)_1(H)_2 + (H)_1(L)_2 + (L)_1(H)_2 + (L)_1(L)_2
\end{equation}

Each of the four terms represents a random observational possibility, with equal probability, of the states of both atoms - they could both be High, Low, or one could be High while the other is Low.  However, consider a situation where we know from physical considerations that the joint state is restricted due to a prior interaction. Perhaps, one being in the state High prevents the other being in the same state High. Then a valid joint QM state is  $(H)_1(L)_2 + (L)_1(H)_2$. However this cannot be written as a product of a state for the particle 1 and another for the particle 2, side by side. There is no state whatsoever for each individual particle that can be represented in QM because  $(H)_1(L)_2 + (L)_1(H)_2$  cannot be written as $(S)_1(S)_2$ where $(S)_1$ and $(S)_2$  represent some (any) state for particle 1 and 2. The essential point is that if particle 1 has some general state, some arbitrary superposition of (H) and (L), and similar state for particle 2, the joint state will always have all combinations, (H)(H), (H)(L), (L)(H) and (L)(L). If one of them was either (H) or (L) then also there will be terms like (H)(H) or (L)(L). There is no way one can get just the two terms (H)(L) and (L)(H) if each particle had any representable physical state in QM. 

\begin{figure}
	\centering
	\includegraphics[width=0.8\linewidth]{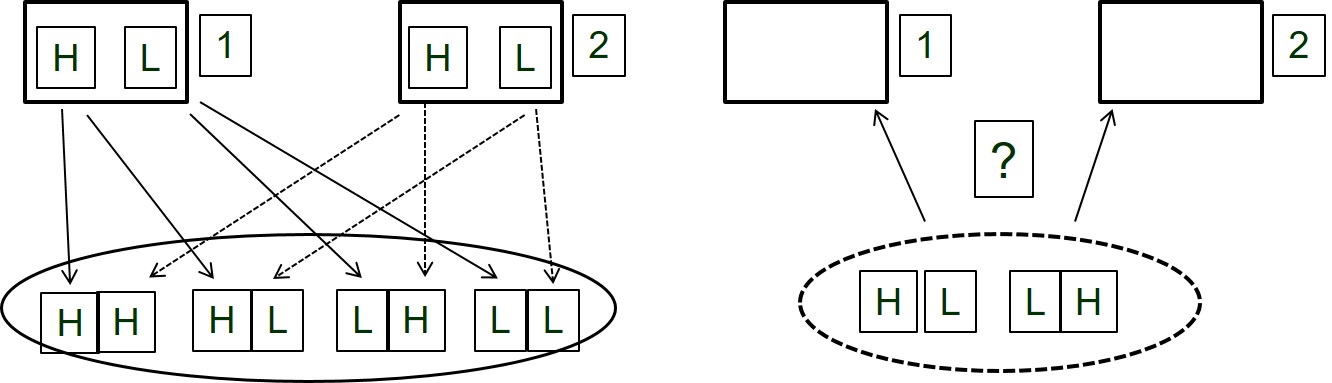}
	\caption{Given states of individual quantum systems, the joint state is a 'product' of both states. However, a joint state does not always factor into product states and in such a situation there is no representation of individual states within quantum theory, in spite of the existence of a joint physical state.}
	\label{fig:figure2}
\end{figure}

The crucial QM ingredient here is superposition, hidden within the algebraic demonstration, which prohibits a familiar model for the source as one that always gives out pairs of particles such that when one particle is in the (H) state the other in the (L) state, each randomly, yet preserving the correlation. Then the only measurement combinations one can have are $(H)_1(L)_2$  or $(L)_1(H)_2$, randomly. Classically such a model is reasonable and that is precisely what was done by proponents of hidden variable theories. However that naive solution falls way too short in many ways \cite{Bell-book,Unni-EPL}. There is a fundamental difference in specifying the physical state as $(H)_1(L)_2 + (L)_1(H)_2$, which is a superposition, and simply as a mixture of pairs of particles with state either as $(H)(L)$  or $(L)(H)$. There are other measurable properties on the same pair of particles with states like (U) and (D) and clever combinations of measurements for different properties have entirely different predictions in a theory with superposition (QM) and in the classical hidden variable theory. That is the content of the celebrated Bell's theorem \cite{Bell-book}. Many experiments were done \cite{Afriat} and the results do not agree with what the classical hidden variable theory expects, obviously. The difficult and disturbing reality of QM superposition is here to stay and an entirely new ways of thinking will be required to understand it, which would a grand intellectual and philosophical relief that many still hope for.

If one asks what the QM state of particle 1 is, with joint state specified as $(H)_1(L)_2 + (L)_1(H)_2$, there is no answer since the possible states of the particle is $(H)$, $(L)$ or in general $(H)+(L)$, and there is no way to factor out any of these from this joint state. A similar conclusion is arrived at for the joint state represented in terms of another physical property as in $(U)_1(D)_2 + (D)_1(U)_2$. Therefore, individual particles in this case have no QM state even though the two together have a valid state. Normally we would have expected that if there are two particles, each should be in some physical state, even if the details are unknown to us in some or all aspects. However, this expectation is broken in the theory of quantum mechanics. This is really surprising since the two particles exist in the material sense, accessible for separate observations with suitable apparatus. Further, the joint state exists, but each particle taken individually has no state at all in the QM representation! Such a state is called an entangled state.  Quoting from Schr\"odinger \cite{Schrod-Camb1935},  who introduced the term,  ``When two systems, of which we know the states by their respective representatives enter into temporary physical interaction due to known forces between them, and when after a time of mutual influence the systems separate again, then they can no longer be described in the same way as before, viz. by endowing each of them with a representative of its own. I would not call that one but rather the characteristic trait of quantum mechanics, the one that enforces its entire departure from classical lines of thought. By the interaction the two representatives (or $\psi$-functions) have become entangled.''

What does this imply for the reality of the atom and the reality of its properties? 
Here one needs to tread very carefully because many accomplished physicists have slipped on the trail. It is perhaps better to state what this state of affairs does not imply in the context of the nature of reality. The first thing to notice is that not being able to represent the physical state of individual particles does not imply that there is no reality to the individual particles themselves, in  a material sense, and this is evident from the fact that the joint description does have components labeled by the identity of each particle. However, within such QM representation each individual particle has no specific state, even an unknown state. Since QM is the only theory known today that accurately describes the statistical features of the microscopic world, with no known exceptions, does this mean also that there could be matter without any physical state of being, not being in any region of space, not being in some unknown direction of spin etc? This is difficult to answer because while common sense abhors such a possibility, the only theoretical framework that we are able to work with to address such questions does explicitly point to that possibility, taken at face value.  

Could this situation then be a limitation of our theory? Could it be that QM is not capable of describing the entirety of reality in the atomic world? Much discussion has taken place along that line as well, starting with the EPR paper. However, the present majority thinking is that QM mechanics is indeed the correct theory with exhaustive scope. If that is the case, the only possibility that remains is to refine our understanding and interpretation within the theory.  The other possibility that the theory is incomplete was argued by Einstein vigorously.   
It seems that all this boils down to understanding what the wave-function is in an ontological sense. It is important to note that the comparison of reality can happen at two different levels in this discussion. One is a comparison between the situation in classical physics and in quantum physics. The other is within QM itself, how the notion of reality as encoded in a representation of physical state gets blurred when one goes from single particle states to multi-particle entangled states.  

How does one explore the reality in the case of two material particles, each individually addressable with experimental apparatus, but neither describable by any individual physical state whatsoever within QM before a measurement? If an observation is made on one of the particles, it does return some result from a possible set for physical quantities like energy or spin, and therefore it acquires a specific representable state within QM.  Since there is a tight correlation of states, knowing the state of one implies knowing the other as well, simultaneously, without even an observation, and then a particle that had no QM state and could be far away from the region of observation also acquires a QM state!  Going back to the specific example of the state  $(H)_1(L)_2 + (L)_1(H)_2$, an observation on the first particle can return the value H or L, randomly, but once a value is observed (say, H), then the other particle necessarily will have the complementary value (L, in this case), even if we do not make an observation. In effect, seeing H on the first particle `collapses' the joint state to just $(H)_1(L)_2$, which is just the individual states of the two particles written side by side. Clearly, each now has acquired a specific state of its own, from a situation of possessing no state, in QM.  The crucial point is that a material entity without any QM representation for its physical state gets a QM state as a result of a measurement on another correlated particle, possibly spatially far away. Note that instead of the physical quantity with values H or L the observer could have chosen to measure on one particle the property that returns values U or D, and then the distant particle would have acquired either the state (U) or the state (D), depending on the outcome of the measurement. Since the physical quantity to be observed in the measurements is free choice and since the exact QM state after the observation depends on this choice, it turns out that the QM state of the distant particle is somehow determined by a free and random choice done far away from the physical system. If we assume that an instantaneous physical influence is impossible and cannot be responsible for this change of state of a particle far away(assumption of Einstein locality), then we have to conclude that QM does not have a faithful and complete representation of the actual physical state of the material particle. That is, after assuming that the  factual physical state of the system (if there is such thing) cannot be influenced from a distance, if the QM representation of the state is indeed influenced by a measurement performed far away (as we discussed), then there is no good correspondence between the physical state and the QM representation. This was the EPR argument of incompleteness of the QM representation, precisely stated. 

In the evolution of the universe where every particle or constituent element shares a history with others, the entanglement and loss of individual reality in the quantum mechanical representation mean that most contents of the universe taken individually have no physical reality at all, within QM. An observation really does not help in improving this situation within the framework of QM because observation itself is an interaction that simply entangles the observer and the observed, dissolving the QM individuality of both!  Therefore, the question why we seem to be able to identify the reality of the individual is an unresolved puzzle with QM. This is same as the problem of emergence of the apparent classical world from the underlying, more fundamental, quantum world.

One may say that the assertion (or belief) that there is some physical state for every physical system, however unknown the details may be, can of course be doubted and questioned. In fact, if this belief is not denied, then QM is already proved to be incomplete by the EPR argument! This is precisely where all the efforts to understand the nature of reality within QM are jammed. However, the majority of physicists go further and believe that QM representation is same as the actual physical state, QM being the complete physics of the situation. Therefore, they believe that the state of affairs one finds in the case of entangled systems implies that there are instantaneous influences (nonlocality) over spatially separated regions that change physical states and their QM representation. This is a serious situation because there is no empirical evidence at all that there is any such a nonlocal influence and the belief is based on trying to translate what happens to QM representations into concepts familiar from classical statistical theories in real space and time. Experiments just measure the correlations, which are larger than what is possible in classical statistical theories with postulated hidden variables and assumed locality. It is possible that such theories can reproduce the experimental results if they are allowed to violate  Einstein locality; if the particles can communicate, they can change their correlation. It is not really physics to propose such telepathic communication between elementary particles. However, suppose one allows that possibility in the hidden variable theories. That has no implication to what happens in quantum mechanics, or nature! But, for some bizarre and distorted reason, people extrapolate from the experimental results that nature is nonlocal, with the instantaneous action at a distance, between the particles.

I may also add that it was a severe misunderstanding of what the Einstein (EPR) argument and the remedy he hoped for were that led to the interest in the local hidden variable theories, starting the 1960s. (The EPR paper did not mention such theories as a possible solution to the incompleteness they discussed). By then, Einstein was not around to clarify his arguments, and guide people away from the wrong track. These theories break explicitly the basic QM features like superposition, and yet hope, naively and ignoring the empirical and logical reasons for QM, to describe the physics of the microscopic world in classical statistical terms. It is straightforward to prove that they grossly violate the fundamental conservation laws \cite{Unni-EPL,Unni-Pramana}. Physicists went to great lengths of experimental effort to rule them out \cite{Afriat}.  Such are the unfortunate consequences of having to use an unobservable in QM whose ontology remains by and large ill-understood. 

\section{From Unobservables to Unspeakables: EPR and Physical Reality}	
A brief discussion clarifying the concept of physical reality in the context of the EPR assertion of incompleteness of quantum mechanics is perhaps appropriate.  My motivation for this discussion is the enormous volume of literature on this subject containing painfully severe misunderstanding of the EPR argument and then equally disturbing assertions about what it implies for reality and locality, the fundamental pillars of classical physics. The misunderstanding can perhaps be ascribed to the historical accident  that the published paper in the American journal Physical Review \cite{EPR}, from which people get these ideas, was written by Podolsky (for reasons of language) and it does not faithfully represent Einstein's essential argument because his `main point was, so to speak, buried in erudition'.  

Indeed, the argument is crystal clear when presented in Einstein's own words. The articles `Physics and reality' \cite{Ein-Franklin} in the Journal of the Franklin Institute (1936) and `Quantum mechanics and reality', written in 1948 for the journal Dialectica \cite{Ein-Dialectica,Born-Ein}, are authentic sources (similar to that in his letters to Schr\"odinger and Popper in 1935, subtle differences notwithstanding).  Einstein considers first a wave-function for one particle that does not specify a sharp position or momentum and asks whether the particle really has a sharp position and momentum `in reality' but the wave-function does not behold that reality, or is that the entire reality is the un-sharp specification that the wave-function represents. The former case implies that the representation is incomplete and also that uncertainty principle does not hold for the reality whereas the wave-function description obeys it. In the latter case, which physicists then and now subscribe to, the realization of a sharp position on observation can be then attributed to the measurement process. Einstein admits that this view alone `does justice in a natural way to the empirical state of affairs expressed in Heisenberg's (uncertainty) principle within the framework of quantum mechanics'.  Then he makes the crucial observation that in this standard view, two wave-functions that differ in more than trivialities describe two different real situations. It is important to notice that Einstein did not bother to define what reality is etc., as done in detail the EPR paper, `burying the main point in erudition'. The next point is the `separation principle', which is also the basis of the principle of Einstein locality.  It is the idea of the independence of existence of objects that are far apart from one another in space. Then external influence on one at location A, like the action of a measurement, has no influence on the other at location B. Then he notices that the QM description of the two-part physical system is in general in terms of the joint wave-function $\psi_{12}$ which cannot be written in terms of wave-functions for the two independent systems: $\psi_{12} \neq \psi_{1}\psi_{2}$, and points out that the wave-functions for the single part systems do not exist at all. The methods of quantum mechanics, however, allow the determination of  $\psi_{2}$ by making a suitable observation on system 1 (S1), without a measurement on system 2 (S2), determining both $\psi_{1}$ and $\psi_{2}$. However, the nature of the resulting $\psi_{2}$, whether it corresponds to a sharp position or a sharp momentum, for example, depends on what measurement is carried out on system 1. Hence, he concludes that according to the choice of measurement on S1, a different real situation is created in regard to S2, because different wave-functions correspond to different realities.  Because we have assumed already that physical reality cannot be altered or created by an act of observation that is spatially separated, the $\psi$- functions cannot be the faithful and complete representation of physical reality.  

That completes Einstein's version of the EPR argument. A reading of the EPR Physical Review paper immediately shows why it does not faithfully represent Einstein's argument and why it is misleading and unnecessarily erudite.

In the section 'Reply to criticisms' in the Schilpp volume \cite{Ein-Schilpp}, Einstein wrote,
\begin{quote}
I close these expositions, which have grown rather lengthy,  concerning the interpretation of quantum theory with the  reproduction of a brief conversation which I had with an important 
theoretical physicist. He: ``I am inclined to believe in  telepathy.'' I: ``This has probably more to do with physics than with psychology.'' He: ``Yes.'' ---
\end{quote}

\section{From Wavefunctions to Quantum Fields: Vacuum vs. Cosmos}
Quantum theory of physical systems consisting of large number of particles led to the concept of a quantum field associated with the particles, a generalization of the wave-function itself.  These fields, one kind for each particle we know of, with the QM version of the electromagnetic field serving as a prototype, are fundamental and essential to modern physics with many success stories of precision calculations. Also, they are considered unavoidable if the quantum theory has to be formulated consistently with the theory of relativity. These fields are as unobservable as any other fields we considered, even though their space-time status is better than that of the wave-function familiar from quantum mechanics of a few particles. One of the fundamental features of such a field theory is that there is an infinite amount of `zero-point' energy (ZPE) in the vacuum state of these fields. Vacuum state is a situation where there are no real particles apparent. Since these fields are modeled after oscillators with all possible frequencies and since an oscillator in QM has a zero point energy (energy in the lowest possible state) worth $hf/2$ where h is the Planck's constant and f the frequency, an infinite or even a large number of oscillators in the field has very large, near-infinite amount of energy. This cannot be avoided, even though all this trouble came about because we had to use some unobservables in the theory.  The early impression was that this really does not matter since a constant background energy has no observable consequence. The general attitude was that what cannot be observed, even an infinite amount of energy, need not bother us. 

However, this relief was short lived because it was realized that there is in fact one dramatic observable consequence of a constant background energy. The rate of expansion of the universe, or the rate at which galaxies move away from each other, is directly proportional to the average energy density in the universe and nothing can be added or subtracted without affecting this observable and measured rate. This expectation from theory (the general theory of relativity) is verified at least approximately even if we do not make new hypothesis about dark matter and so on. Even though the amount of matter that is visible through the light emitted is estimated to be only 5\% of what is required to explain the observed rate of expansion, we may say that there is an approximate consistency. In contrast,  the vacuum zero-point energy calculated in the context of quantum field theories far exceeds, by numbers beyond imagination, what is reasonably consistent with the measured expansion rate of the universe. This is vexing problem of the large cosmological constant. Thus, observational cosmology provides a powerful test of our theories and the verdict goes against the theoretical structure (of quantum fields) we use today. No solution has been found and a solution may well have to deny the physical reality of quantum fields. This will also mean that we will have to reconstruct the relevant physical theories in terms of quantum fluctuations and energy of only matter, without the zero-point energy of the quantum fields.  At present it is not known whether this is possible even in principle.    

The Casimir force, Lamb shift, and the spontaneous emission from excited atoms are phenomena cited in support of the physical reality of the wave modes of the quantum fields in their `vacuum state', when there are no real particles. However, the Casimir force, the prime example, can be derived as the interaction of the quantum fluctuating dipoles (atoms) in the material making the two surfaces, as the retarded van der Waals force, or as the integrated Casimir-Polder force between a material boundary and an atom \cite{Milonni}. Since all real boundaries are equal to the factual presence of matter with quantum zero-point motion, and not merely static mathematical conditions, the necessity of the vacuum modes in the Casimir force cannot be insisted because that would be double-counting; either picture is mathematically consistent when invoked alone. However, since matter and its zero point fluctuations are the only reality that is directly observed and verified, the mode picture can be seen only as a calculational tool, without physical reality to the wave modes. This argument is logically robust and it solves the problem of the divergent ZPE, since the matter density in the universe is finite and its ZPE is negligible.

Some aspects of  the `successful' standard model of particle physics, which depends on what is called '`gauge freedom' or freedom for changing the values of certain field quantities in the theory without observable effects, also may have to be reconsidered in detail in the context of the consequences for the dynamics of the universe. 

\section{Postulates as Reality}
Postulates in physical theories are formal general statements on fundamental physical phenomena, usually and preferably extracted from observational evidence. What is peculiar in physics is that lack of falsifying evidence also can be taken as supporting evidence for a postulate, based on which the theory is constructed.  If the theory turns out to be consistent with observations, especially those that come after the construction of the theory, the faith in the postulate is strengthened. Since accessible and realizable observations and experiments are vastly limited in comparison to the scope of a general theory, this situation allows postulates that are familiar and not easily falsifiable to be projected as truth and reality. The ill effects of this scenario is more and more evident in present day physics with many postulates, and several unobservable entities, outside the immediate reach of experimentation. 
  
What will be surprising is that even the most familiar and all-pervading of fundamental physical theories -- the special theory of relativity -- suffers from this problem. Its fundamental postulate of the absolute constancy of the velocity of light relative to inertial observers  is neither proved nor contested, but most physicists believe that it is amply verified to be true. The real nature of the physics of light and relativity is intimately related to the universe and matter and therefore we turn to a discussion of a new and essential paradigm of physics in relation to the universe in which it is formulated, tested and applied.

\section{Universe, Space, Time and Notions of Reality}
All our fundamental theories, without exception, were constructed and completed before we acquired any significant knowledge about the universe - its matter content, extent, history and dynamics. There are serious philosophical and physical issues involved here because all our theories assume, and in some cases require, empty space as their backdrop. However, matter-filled universe is also very gravitational and since we know now that gravitational interaction can change the rate of clocks and the length standards, the entire edifice of metrical physics needs a serious reconsideration. 

To start with, all our theories are conceived, completed and tested in the presence of all the matter in the universe and if the relevant gravitational effects are not included and accounted for systematically, we could be working with deficient or incomplete theories, even when there is apparent agreement between what we expect from a theory and what we see observationally. All experimental results naturally include any cosmic gravitational effect that may be there, since all experiments are done in the unavoidable presence of such matter, whereas the theories presently used to interpret them assume an empty space-time as their background. A most plausible scenario is that many physical effects we see are in fact gravitational effects linked to the matter in the universe whereas our theories describe them as due to other reasons, sometimes dubious. This indeed is the case is now evident from several considerations but we will discuss only one or two relevant points here since our focus in this section is on examining the status of the unobservable space as a physical entity in physical theories. The full program to formulate physics in the `once given universe', called Cosmic Relativity has many interesting features and predictions for relativity effects, electrodynamics and propagation of light.  \cite{Unni-Cosrel,Unni-PIRT2,Unni-ISI}. Its cardinal prediction of the Galilean nature of the propagation light is now verified in experiments \cite{Unni-ISI,Unni-TPU}.

Physicists seem very proud of their achievement of banishing the old `ether' from physics and this is usually highlighted as an example of removing unobservables from physical theory. In fact, there are really no good empirical reasons that are brought forth in support of this achievement and the typical reasoning quotes the success of the special theory relativity that rendered the ether irrelevant. Treated rigorously from an empirical point of view, an unobservable like ether could never be `disproved'. Often quoted remarks that the null result of the Michelson-Morley (M-M) experiment and its variants disprove the reality of ether are based on superficial understanding of both the physical analysis of the experimental result and the history of the issues involved. In fact, what happened was what Poincar\'e predicted. In `Science and Hypothesis', he wrote \cite{Poin-Hypo},  
\begin{quote}
	Whether the ether exists or not matters little -- let us leave that to the metaphysicians; what is essential for us is, that everything happens as if it existed, and that this hypothesis is found to be suitable for the explanation of phenomena. After all, have we any other reason for believing in the existence of material objects? That, too, is only a convenient hypothesis; only, it will never cease to be so, while some day, no doubt, the ether will be thrown aside as useless. 
\end{quote} Special relativity, however, certainly deviated from the premise `everything happens as if the ether existed'.  Einstein's hypothesis of constant relative velocity for light is an elegant solution of the null result of the M-M experiment, and the Lorentz-Fitzgerald contraction hypothesis is another valid solution. Ironically, many  experimenters continued to write that they were `proving' the relativistic effects (of length contraction and time dilation) through their null results, well into the 1930s. Since the experimenters were in the very laboratory frame in which the interferometry experiments were performed, special relativity denies any length contraction or time dilation effects whereas many experimenters preferred to depend on such physical effects, rather than special relativity, to explain their null results. 

Success of the special theory of relativity brought empty space and vacuum as major physical entities into physics. Not realizing that there was no empty space in reality and that the entire universe was filled with matter in some form, allowed a misguided enthusiasm about bestowing reality status to empty space and an associated time, together called space-time, even though there is no way to think about these without matter. Gravity was interpreted as the geometrical distortions that happen to space-time due the presence of matter or by a prior design. No doubt, the result was an elegant theory with great success but it alienated the physics of gravity from the physics of other interactions which could not be formulated the same way. Fundamental to these developments was the equivalence principle which is the universal relation between  the `charge' of the gravitational interaction, and `inertia' in dynamics; both were identified and called by the same name, `mass'. In this course, Ernest Mach's brilliant insights \cite{Mach} on the possible origin of  inertia as due to interaction with matter around was dismissed as irrelevant, in spite of a hint of evidence in favour, in general relativistic effects related to rotation \cite{Ciufolini}.  Later, this was distorted, by misinterpreting certain experimental results, to mean that what Mach said was incorrect \cite{Unni-Mach}. Throwing out the unobservable ether is one thing, but daring to throw out the readily observable matter-filled Universe from consideration is a colossal error of judgment. In any case, the general theory of relativity is considered our best theory, and it is well tested. However, the aura about its geometrical nature with stress on empty space and time as physical entities is in fact overstated. This could very well be the red herring that delays the goal of a quantum theory of gravity \cite{Unni-Sen-QG}.  It is even likely that the geometrical interpretation of its application to the dynamics of the universe could contain some features that will need revision in future.

\section{Special Relativity: The Theory Built on Several Unobservable Pillars}
It is interesting to note that even real physical length contraction and time dilation are unobservables for the observer moving with the scale and clock because all such physical measurements are comparisons and the standard scale and clock used for the comparison also suffer exactly the same contraction and dilation. This unobservability for a comoving observer is of course different from the universal unobservability of some theoretical entities (`free creations of the human mind') we discussed. It is precisely the unobservability of these real physical effects that allows the special relativistic assertion that there is in fact no contraction or dilation in the comoving frame. Once the contraction is denied, then and only then, the invariance of the relative speed of light can be postulated as the explanation of the null results in the experiments.

This discussion shows how easily one can be mislead about the reality and truth of phenomena in physical nature because of our equating reality to what could be observed.  Realizing that a measurement is often a comparison with a physical standard that is subject to the same physical effects we are trying to measure is of utmost importance in evaluating and testing a theory with empirical and logical rigour.  Assertions that the standard tests of the M-M type prove special relativity and the constancy of the relative velocity of light lack this rigour \cite{Brown}.\footnote{It is elementary to prove that neither the two-way experiments like the M-M experiment or the Kennedy-Thorndike experiment, nor any one-way experiment with spatially separated clocks can demarcate between Galilean propagation of light ($c\pm v$), as in the Lorentz-Poincar\'e ether relativity, and the invariant relative velocity of light in Einstein's special relativity. All claims to the contrary are incorrect.} In fact, as soon as round trip comparisons are included as possible tests of relativity theories, in which case cumulative effects like time dilation become observable, the weakness of special relativity is revealed \cite{Unni-ISI,Unni-Twin}.  Clearly, `unobservable' in physics does not always imply `unreal'.  Reality cannot be denied just because the phenomena is not  observable by a limited and constrained class of observers. 

The special theory of relativity implements the principle of relativity - the invariance of the laws of physics in all inertial frames - by postulating the absolute invariance of the relative speed of light in all frames in empty space. Its supporting pillars are two unobservables - empty space (and its velocity independent isotropy) and the absolute invariance of the relative one-way speed of light. Einstein realized that the one way speed of light over a stretch of spatial distance, which requires two clocks, is an unobservable in a limited sense. Since this measurement requires synchronization, the one-way speed depends on the convention for synchronization. Though only the two-way speed of light is experimentally shown to be an invariant, it is possible to postulate that even the one-way speed is an invariant by suitably defining a synchronization, embodied in the first order term of the Lorentz transformation. In fact, if the time light takes to cover a distance $\Delta x$ is $\Delta t = \Delta x/c$ in a stationary frame, it should take an additional duration of $v\Delta t/c = v\Delta x/c^2$ relative to an observer moving at velocity $v$, if the relative velocity of light is not a constant. 

\begin{figure}
	\centering
	\includegraphics[width=0.4\linewidth]{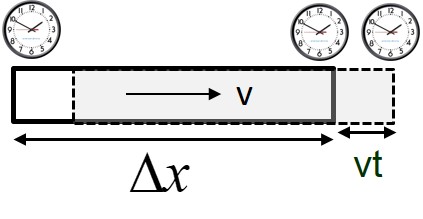}
	\caption{Making Galilean light Lorentzian relative to a moving frame by a direction and distance-dependent adjustment of time $\delta t=v\Delta x/c^2$, of a clock that is $\Delta x$ away.}
	\label{fig:figure3}
\end{figure}

However, if we postulate that the clock at the distance $\Delta x$ requires a distance-dependent adjustment for some reason and decree that it is to be adjusted back by the amount  $\delta t= v\Delta x/c^2$, the additional duration that (Galilean) light takes is cancelled. Then the apparent duration is just $\Delta t = \Delta x /c$, which is the same as the duration relative to observer at rest (figure 3).  This is exactly like the Galilean adjustment of the moving coordinate reference X' where the coordinate values are compensated for the movement at velocity v for duration t as $x′ = x-vt$. But this adjustment of the clock allows us to claim that the speed of light measured with such a (fake) clock remains invariant to first order in v/c! Herein lies the circularity of the reality created by the special theory of relativity: propagation of light is used to define time of separated clocks and the same time is used to claim the invariance of the relative speed of light. (It is also relevant to note that the first order `correction factor' is exactly what an absolute frame theory would give from its second order effect for the time shown by a clock synchronized at one end of the scale  and then slowly moved to the other end, $\Delta x$ distance away, when the entire system is moving at speed v.)  It is important to stress the factual situation that no experiment has ever shown, independent of conventions or circular arguments, that the one-way speed of light is an invariant constant. (The smaller second order effect, $(1-v^2/c^2)^{1/2}$,  is in fact the only empirically verified correction to moving clocks, after round trips).  In fact, the moment we can identify a universally synchronized clock -- any phenomenon that serves as a time-keeping device which is the same everywhere in an extended spatial region - the hypothesis of the invariance of the one-way speed of light can be tested. Such a `Galilean clock' will also be in conflict with the Lorentz transformation because it will show that the clock adjustment $\delta t= v\Delta x /c^2$ at distance $\Delta x$ is not real time, but it is unavoidable because inertial motion cannot be detected.  This identifies exactly where the weak point of the theory lies.

\section{Relativity and the Universe}
A serious conflict arises between what we know about the universe today and what was constructed as the `correct' theory of relativity in 1905. The special theory of relativity described all relativistic effects as due to relative motion between different inertial observers and maintained that the notion of special preferred frame, or an absolute frame of reference, cannot be sustained. In the special theory, the validity of the principle of relativity (the impossibility of detecting uniform motion by any physical phenomena) was linked to the invariance of the empty space - empty space remains homogeneous and isotropic to all inertial observers, independent of their velocity, and all these observers are hence equivalent. Then there is no universal time, nor a universal reference that makes the concept of absolute rest meaningful. Indeed, just the requirement that the geometry (metric) of empty space remains invariant under motion leads to the Lorentz transformations that form the entire basis of the theory. However, real space as we observationally know today is not empty at all. It is filled with matter. Therefore, the space appears very different to a moving observer in comparison to an observer who is stationary relative to the average matter distribution.  In relation to a moving observer, the entire matter moves as a large directional current and the space appears anisotropic, proportional to the velocity. This is in fact easily observable by a moving observer as the `dipole anisotropy' of the cosmic microwave background radiation (CMBR). There is indeed operational and conceptual meaning to a state of absolute rest in such a situation. Since space does not remain isotropic in a moving frame, the underlying geometry is clearly anisotropic in moving frames and the inconsistency of the Lorentz transformations to describe the situation  is immediately evident - Lorentz transformations preserve isotropy and  homogeneity whereas isotropy is broken under motion in real matter filled space. There is even a universal absolute time since the evolving matter and radiation that define time have essentially the same history everywhere in the universe. Operationally, the slowly decreasing temperature of the cosmic microwave background radiation can be taken as this universal time and it is automatically synchronized everywhere in the universe. Thus, we have identified a reasonably precise and useful Galilean clock \cite{Unni-ISI}. Indeed, the whole premise of special relativity is made invalid by the evolving observable matter-filled universe. Replacing the unobservable ether and the even more unobservable empty space with an observable matter-filled universe as the background arena of physics restores logical and empirical consistency of physical theory with the gravitational presence of the universe. 

One can go further and answer several questions that arise. The anisotropic space and its geometry in the frame of the moving observer can be correctly described by the Galilean transformation and, along with the gravitational action of the matter and matter current in the universe, all observed relativistic effects like time dilation are explained \cite{Unni-ISI}. This framework also shows that inertia is indeed the resistance to acceleration arising in the gravitational interaction with the entire matter in the universe, as Mach had guessed. Most surprisingly, Newton's law of motion emerges naturally as a gravitational effect on motion \cite{Unni-Bern}.  Along the way, the principle of relativity and the equivalence principle, hitherto considered as essential postulates for physics in the unobservable empty space materialize as consequences of the gravitational action of the matter-filled universe. The lesson from this development is that relativity without cosmic matter is empty relativity with no physical effects on moving clocks etc. Physics in empty space is meaningless. Matter and its gravity indeed are the real basis for relativity and dynamics. With this realization, and several empirical consequences and supporting evidences in the context of detailed behaviour of clocks and the one-way propagation of light, the concept of absolute rest, absolute space and time are firmly back into physics and the issues of reality are linked inseparably to the reality of matter and its gravitational action. The most drastic consequence of the return of the absolute is understandably in the behaviour of light. Unambiguous empirical evidence for the first order non-invariance of the relative velocity of light \cite{Unni-ISI}, while preserving the upper limit of c for all motion in the cosmic frame, restores the underlying reality of a cosmic matter-based paradigm of relativity and dynamics. That we continue to use the unobservable gravitational field for a convenient description of this scenario is another matter! 

\section{Concluding Comments}
It seems that the use of unobservable entities is unavoidable for theorizing about the physical nature. However, it is very important to be watchful and critical about their overall consistency within the structure of physics as well as their implications to all conceivable physical situations. The problem is not really the use of unobservables, but the insistence on non-critically extrapolating from the resulting successes of the physical theory on a limited set of predictions to the unquestionable reality of the unobservable. Unobservable wave-functions and quantum fields in modern physics emerged naturally from theoretical needs, but their ontological status is a source of constant debate and worry. While the wave-function can be lived with, but for the bitter feeling and frustration of not being satisfied with sufficient understanding of the space-time picture of cause-effect relations in the microscopic world, the damaging inconsistency of some features of the quantum fields with the dynamics of the universe is something that needs cure in a future theory.  It is clear that the inseparable and ever-present gravitational link to the matter-filled universe is in conflict with a theoretical framework that relies on the unobservable empty space and the associated time as its basis of physical effects. In fact, physical changes that have no cause in physical interactions are suspect and they always point to the need for a new theory. At least in this case, a much better realistic basis is provided by the paradigm of Cosmic Relativity that derives all relativistic physical effects as due to matter and its gravity in the universe. With this realization, the absolute frames of space and time are back in physics. Not surprisingly, this paradigm with its universal cosmic links and physical inseparability also holds human concepts of harmony and beauty \cite{Unni-Leonardo}.  

While a program to banish all unobservables from physical theory is unlikely to succeed, and perhaps even unnecessary, they are both the strong and weak points of theoretical descriptions of physical nature.  Analyzing them for empirical and philosophical consistency and integrity is always a promising path towards a better theory.

\section*{Acknowledgments} 
I thank Partha Ghose and Peter deSouza for the invitation as well as for the friendly and warm hospitality at the Indian Institute of Advanced Studies, Shimla, during the superbly nourishing seminar, `The Nature of Reality: The Perennial Debate' (March, 2012). Discussions with several participants, especially the philosophers who expect better rigour, have spurred and helped me to sharpen some of the thoughts presented in this paper. Conversations with Martine Armand on the contents and style have helped in a clearer presentation.

\end{document}